\documentstyle[preprint,aps]{revtex}

\begin{document}

\draft
\title{Specific Heat of defects in the Haldane System Y$_2$BaNiO$_5$}
\author{K. Hallberg, C.D. Batista and A.A.Aligia }
\address{Centro At\'omico Bariloche and Instituto Balseiro}
\address{Comisi\'on Nacional de Energ{\'\i}a At\'omica}
\address{8400 S.C. de Bariloche, Argentina.}
\date{Received \today }
\maketitle

\begin{abstract}
We calculate the specific heat of the antiferromagnetic spin-1 chain compound 
Y$_2$BaNi$_{1-x}$Zn$_x$O$_5$  in the presence
of a magnetic field. The low-energy spectrum of a Heisenberg Hamiltonian, which
includes realistic anisotropies, has been solved
using the density matrix renormalization group. The observed Schottky anomaly
is very well described by this theory. For large chains ($N>50$), contrary to previous
 interpretations of the specific heat data, we find 
$S=1/2$ states. These results are thus also consistent with
 electron-spin-resonance data for NENP. 
\end{abstract}



\narrowtext

A great deal of interest in one dimensional Heisenberg chains with
nearest-neighbor antiferromagnetic (AF) exchange coupling, $J$, has been
originated by Haldane's conjecture that integer-valued spin chains would
exhibit a gap in the spin-wave excitation spectrum \cite{hal}, in contrast
to half-integer spin chains which would be gapless \cite{bet} with a linear
dispersion relation above the ground state. This quantum many-body phenomena
is different from the usual source of gaps in magnets, namely, single-ion
anisotropy, which does not involve correlation effects. Subsequent
observations of large gaps in spin-1 quasi-one-dimensional systems present
in the structure of some compounds like CsNiCl$_{3}$ \cite{buy}, [Ni(C$_{2}$%
HgN$_{2}$)$_{2}$(NO$_{2}$)]ClO$_{4}$ (NENP)\cite{ren,kat}, and Y$_{2}$BaNiO$%
_{5}$ \cite{che,dar}, have confirmed Haldane's observation.

Affleck {\it et al. }\cite{aff} have shown that the exact
ground state of the Hamiltonian 
$\sum_{i}{\bf S}_i\cdot {\bf S}_{i+1}+({\bf S}_i\cdot {\bf S}_{i+1})^2/3$ is a
valence-bond-solid (VBS) state. In this state, each
$S=$1 spin is represented by symmetrization of two $S=1/2$
entities. These $S=1/2$ spins at each site are coupled with nearest-neighbor 
$S=1/2$ spins, one to the left and the other to the right, to form singlets.
Hence, an open chain has two unpaired $S=1/2$ spins, one at each end. 
They \cite{aff} have also proposed that the VBS state describes
qualitatively the physics of the Heisenberg model (without the
biquadratic term). In
agreement with this proposal, exact diagonalization \cite{ken} of finite open
chains shows that the four lowest-lying states are a triplet and singlet
whose energy separation approaches zero exponentially with increasing
length. Monte Carlo \cite{miy} and density-matrix renormalization-group
(DMRG) \cite{whi} studies clearly show the presence of $S=1/2$ end states.
This picture is also supported by electron paramagnetic resonance (EPR)
measurements of NENP\ doped with non-magnetic ions \cite{gla}, where
resonances corresponding to the fractional spin $S=1/2$ states at the
``open'' ends of the Ni chains were observed.

However, Ramirez {\it et al.} \cite{ram} also tested the presence of free $%
S=1/2$ states by studying the specific heat of defects in NENP\ and Y$_{2}$%
BaNiO$_{5}$, with magnetic fields up to 9T and temperatures down to 0.2K. On
the basis of a comparison between two very simplified pictures, they found
that the shape and magnitude of the Schottky anomaly associated with the
defects in Y$_{2}$BaNi$_{1-x}$Zn$_{x}$O$_{5}$ are well described by a simple
model involving spin-1 excitations, instead of the $S=1/2$ excitations of
the VBS. These results are in contrast with the above mentioned EPR
measurements of NENP \cite{gla} .

While previous theoretical 
studies of the spin-1 Heisenberg chain \cite{ken,miy,whi}
confirm the existence of free $S=1/2$ spins at the ends of sufficiently long
chains in agreement with the EPR measurements, it seems difficult to
reconcile these results with the specific heat measurements and no
theoretical calculation of the latter exists so far.

In this Letter, we solve the low energy spectrum of the appropriate
Hamiltonian which describes the Ni chains of Y$_{2}$BaNiO$_{5}$ using DMRG
and finite size scaling. Assuming that defects are randomly distributed
(which leads to a Poisson distribution for the length of the open chains),
we can calculate the specific heat for any concentration of non-magnetic
impurities. Our results show that there is no contradiction between EPR \cite
{gla}\ and specific heat measurements in Y$_{2}$BaNi$_{1-x}$Zn$_{x}$O$_{5}$ 
\cite{ram}. They confirm the existence of spin-$\frac{1}{2}$ excitations for
sufficiently long chains and are in excellent agreement with the curves
measured by Ramirez {\it et al.} \cite{ram}.

Y$_{2}$BaNiO$_{5}$ has an orthorhombic crystal structure with the Ni$^{2+}$ (%
$S=$1) ions arranged in linear chains with a nearest-neighbor AF
superexchange coupling. The interchain coupling is more than three orders of
magnitude weaker, making this compound an ideal one-dimensional
antiferromagnetic chain. While each Ni atom is surrounded by six O
atoms in near
octahedron coordination, the true site symmetry is $D_{2h}$, and the
appropriate Hamiltonian reads \cite{sak,xu,gol}:

\[
H=\sum_{i}\{J{\bf S}_{i}\cdot {\bf S}%
_{i+1}+D(S_{i}^{z})^{2}+E[(S_{i}^{x})^{2}-(S_{i}^{y})^{2}]\}-g\mu _{B}{\bf %
B\cdot S}_{tot} 
\]
where $z$ is along the chain axis and ${\bf S}_{tot}$ is the total spin.
Recent estimates based on fits of the Haldane gaps (in $x,\;y,\;$and $z$
directions) measured by inelastic neutron scattering , indicate $J\sim
280K\; $, $D\sim -.038J\;$, and $E\sim -.013J\;$\cite{sak,xu}.

As pure Y$_{2}$BaNiO$_{5}$ has a Haldane gap $\sim 100K$ \cite{dar,sak,xu}, the
spin wave contribution to the specific heat $C(T)$ is negligible below 7$K$.
In this temperature range, $C(T)$ is dominated by the effect of the defects,
which is manifested as a $1/T^2$ rise in $C(T),$ and a Schottky anomaly
in the presence of an applied magnetic field. The height and width of the
Schottky anomaly strongly depends on the spin value of the low energy
excitations. For these reasons, it is necessary to solve the low energy
spectra ($\omega \leq 10K$) of $H$ for all values of $N$ to have an accurate
theoretical picture.

By means of the DMRG method, we have calculated the two lower eigenenergies
in the $S_{tot}^{z}=0$ subspace and the lower one with $S_{tot}^{z}=1$ (this
state is degenerated with the $S_{tot}^{z}=-1$ due to the time-reversal
symmetry of $H$) for all $N\leq 40$ and $E=0.$ The energy difference between
the excited states and the ground state decays exponentially to zero with
increasing $N$. This behavior, shown in Fig. 1 for an arbitrary value of
anisotropy $D=-0.1$, allows us to extrapolate the energies to all values of $%
N>40$, and demonstrate that the free $S=$1/2 spins at the end of the chain
persist in presence of anisotropy. This issue is easy to understand
considering that $S=$1/2 spins cannot be affected by anisotropy due to
time-reversal symmetry. At this level, it is important to remark that the
two $S=$1/2 spin excitations, predicted by VBS, have a finite localization
length $l\sim $ 6 sites \cite{ken,miy,whi}. Therefore, while they are nearly
free for large ($N>>l$) open chains, the interaction between them is
considerable when the length of the chain $N$ is comparable to $2l.$ This
interaction splits the two $S=$1/2 states into a singlet (ground state for
even $N$) and a triplet (ground state for odd $N$) one \cite{lie}.

The difference between any two energies of the above mentioned low-energy
states is linear in $D$, and the quadratic corrections are negligible \cite
{gol}. This widely justifies the validity of perturbation theory to first
order in $D$. By symmetry then we can include the term $%
\sum_{i}E[(S_{i}^{x})^{2}-(S_{i}^{y})^{2}]$ to first order. Thus we find the
following low-energy effective Hamiltonian including a spin $S=1$ and a
singlet state $\left| 0\right\rangle $:

\begin{eqnarray*}
H_{eff} &=&(E_{0}(N)+\alpha (N)D)\left| 0\right\rangle \left\langle 0\right|
+D\beta (N)S_{z}^{2} \\
&&+E\beta (N)(S_{x}^{2}-S_{y}^{2})-g\mu _{B}{\bf B\cdot S}
\end{eqnarray*}
where $E_{0}(N),\;\alpha (N),\;$and $\beta (N)$ are functions of the chain
length $N$ (determined from the DMRG data). The validity of the last term
has been verified explicity by calculating the matrix elements of $%
S_{tot}^{+}$, and $S_{tot}^{-}$ for all chains. $H_{eff}$ determines the
thermodynamics of the system at temperatures well below the Haldane gap.

For a random distribution of defects, the specific heat per elementary cell
of Zn-doped Y$_{2}$BaNiO$_{5}$ is:

\[
C(B,T)=\sum_{N=1}^{\infty }\;x^{2}(1-x)^{N}\;C_{N}(B,T) 
\]
where $C_{N}(B,T)$ is the specific heat of a segment of length $N$ described
by $H_{eff}$, and $x$ is the concentration of missing Ni atoms.

In order to minimize the effect of the lattice on $C(T)$, Ramirez {\it et
al. }\cite{ram} plot the data, measured in Y$_{2}$BaNi$_{0.96}$Zn$_{0.04}$O$%
_{5}$ , as the difference $[C(3$T$)-C(6$T$)]/T$ . In Fig. 2, we compare this
difference calculated with $H_{eff},$ with the data taken from Ref. \cite
{ram}. As the samples were
powdered in order to ensure good thermal equilibration \cite{ram}, the
theoretical curve was obtained by averaging over all magnetic field
directions. We have taken $J=280K$ , $D=-0.038J$ consistent with neutron
scattering experiments \cite{sak,xu} and $E=-0.032J,$ somewhat larger than
the experimental value. We have also considered $g=2.35$ which is a
reasonable value for the $g$ factor \cite{gla,bal}. The agreement between
experiment and theory is excellent. Note that although the fit with the
singlet-triplet model is good, the result we obtain here is better,
particulary for temperatures around $0.5K$, and between $3$ and $6K$. We
also used a doping concentration of $x=0.0428$ higher than the nominal value 
$x=0.04$. This is due to the existence of native defects, also present in
the pure compound. By fitting the data corresponding to pure Y$_{2}$BaNiO$%
_{5}$ we obtain $x=0.0028$ instead of the value $x=0.008$ obtained 
with the singlet-triplet model in Ref.  \cite{ram} (see Fig. 3).
Note that for small values of $x$, the height of the peak is
proportional to $x$.

 From Figs. 2 and 3 it is clear that the inclusion of interactions between end 
$S=$1/2 spins through the chain via the full Hamiltonian $H$, together with
anisotropy, eliminates the apparent discrepancy between VBS theory
predictions and measured specific heat \cite{ram}. In Fig. 4 we have
separated the contributions of different chain lengths to $[C(3$T$)-C(6$T$%
)]/T$ . The full line corresponds to the result for non-interacting $S=$1/2
end spins. This behavior is obtained for $N\gtrsim 50$ where the
singlet-triplet gap is much smaller than $g\mu _{B}B$. The effect of
anisotropy is stronger for shorter chains, changing the position and height
of the Schottky peak. This behavior is mainly due to odd chains because they
have an $S=$1 ground state. The anisotropy splits the $S_{z}=0$ and $%
S_{z}=\pm 1$ (ground state for $D<0$ ) triplet states, and the energy
separation $\Delta (N)$ decreases with $N$ (as can be inferred from Fig. 1).
As one can see from Fig. 4, for large chains ($N>30$), the result approaches
asymptotically the VBS prediction, {\it i.e. }the peak is higher and shifted
to slightly smaller temperatures as compared to the experimental result.
However, the curves for shorter chains ($%
\Delta (N)\gtrsim g\mu _{B}B$) show the opposite behavior, and both
tendencies are compensated in the final result. To understand this effect
let's consider the simplest case of $E=B_{x}=B_{y}=0$ and odd $N\;$. The
energy difference corresponding to the Schottky anomaly, $E(S_{z}=0)-$ $%
E(S_{z}=-1)$, becomes $g\mu _{B}B_{z}+\Delta (N)$ when a magnetic field is
applied. This explains the shift to the right registered for the shorter
chains. As a consequence, the sum over small ($N\lesssim 20$) values of $N$,
weighted with the Poisson distribution, yields a Schottky peak of larger
width and lower height .
The anomalous features at low temperature in Figs. 2 and 4 are due to
the contributions of chains with even $N$ for which the difference
between the energy of the triplets with $S_z=\pm 1$ and the singlet
ground state becomes of the order of the Zeeman energy $g\mu_BB$.
For $6T$ this crossing occurs for $N$ between 18 and 20 giving rise to
the large negative contribution for the corresponding curve
in Fig. 4. For $B=3T$ the crossing occurs for $N$ between 22 and 24
(see the $21\le N \le 30$ curve in the same figure).

Our results allow us to understand why the $S=1$ states of short chains were
not seen \cite{note} in the EPR experiments in NENP\cite{gla}. For  $D/J\sim
0.2$  the splitting $\Delta (N)$ is greater or of the order of the applied
magnetic field $g\mu _{B}B_{a}$ ($B_{a}\sim 0.35$T ) when $N<38$, while for $%
N>54,$ $\Delta (N)<0.1g\mu _{B}B_{a}$ . These larger chains ($76\%$ of the
total number of chains for $x=0.005$ ) contribute to the same EPR signal
with total weight $0.0038$ per transition metal atom. On the other hand,
each of the chains with $N<38$ ( weight $\lesssim $ $.0002$), give place to
a different gyromagnetic frequency larger than the one they considered \cite
{gla}.

In conclusion, by solving the low-energy spectrum of a Heisenberg
Hamiltonian $H$ which includes axial and planar anisotropy we have
reproduced the low-temperature and high-field specific-heat data measured in
Y$_{2}$BaNi$_{1-x}$Zn$_{x}$O$_{5}$ \cite{ram}$.$ Our results are consistent
with valence-bond-solid predictions of $S=$1/2 end chain excitations,
which are asymptotically free for large chain segments. However, their
interaction for short segments are critical for the understanding of
the specific heat behavior.
These results remove the apparent discrepancy between the specific heat data
for Y$_{2}$BaNiO$_{5}$, interpreted in Ref.\cite{ram} with a singlet-triplet
model, and electron paramagnetic resonance data for NENP.

We would like to thank E. R. Gagliano for useful discussions.
K.H. and C.D.B. are supported by Consejo Nacional de Investigaciones
Cientif\'{\i }cas y T\'{e}cnicas (CONICET), Argentina. A.A.A. is partially
supported by CONICET.

\newpage\ 

Figure 1: Difference between the energy of the$\;$triplet $S_{z}=\pm 1$ 
($S_{z}=0$) and the singlet states  in the presence of anisotropy $D$ are
denoted with circles (squares) . a) $N$ even, b) $N$ odd. The full line is an
exponential fit of these differences for $N\geq 19$.

Figure 2: Specific heat difference between 3 and $6T$ for
Y$_{2}$BaNi$_{0.96}$Zn$_{0.04}$O$_{5}$ \cite{ram}. 
The dashed and solid lines correspond to fits
with the prediction of $H$ $(J=280K,\;D=-0.038J,\;E=-.032J,\;x=0.0428,
\;g=2.35)\;$ and the singlet-triplet model $(x=0.048)$.

Figure 3: Specific heat difference between 3 and $6T$ for undoped 
Y$_{2}$BaNiO$_{5}.$ Solid circles are experimental data \cite{ram}. 
The solid 
line corresponds to a fit with the prediction of $H$ for the same parameters
as in Fig. 2 except for $\;x=0.0028,\;$and the dashed line to the 
singlet-triplet model with $x=0.008$.

Figure 4: Contributions of chains in different length intervals to the
specific heat difference between 3 and 6T calculated with $H$ for the same
parameters as Fig. 2. The full line corresponds to the prediction of VBS
theory for non-interacting $S=$1/2 moments. The final result is obtained
adding each curve with the weight denoted inside the figure.


\begin{references}
\bibitem{note}  The $S=1$ spins (as opposed to $S=1/2$) are affected by
anisotropy and to first order the splittings generated by an applied
magnetic field are independent of microwave frecuency contrary to what is
observed \cite{gla}.
\end{references}
\end{document}